\documentclass{article}

\usepackage[labelfont=bf,font=small]{caption}
\usepackage{graphicx}%
\usepackage[utf8]{inputenc}       
\usepackage{amsmath}
\usepackage{authblk}          
\usepackage{amsfonts}
\usepackage{amssymb}
\usepackage[numbers, sort&compress]{natbib}  %
\usepackage{bibunits}
\usepackage[table, dvipsnames]{xcolor}
\usepackage{lineno}            
\usepackage{xspace}
\newcommand{\um}{\ensuremath{\mu\mathrm{m}}\xspace} 
\usepackage[strict]{changepage}   
\usepackage{nicefrac}
\usepackage{lipsum}
\usepackage{paralist}
\usepackage{siunitx}
\usepackage{pifont}
\usepackage{tikz}
\newcommand*\circled[1]{\tikz[baseline=(char.base)]{
            \node[shape=circle,draw,inner sep=1pt] (char) {#1};}}

\defaultbibliographystyle{naturemag}
\defaultbibliography{references}

\usepackage{soul}               
\usepackage{xcolor}
\setstcolor{red}
\setul{}{1.5pt}

\title{Non-reciprocal electrooptic intermodal scattering with momentum engineered RF waves}

\author{
    Jieun Yim$^{1*}$, Gwan In Kim$^2$, Violet Workman$^3$, Seho Kim$^2$, Omar A. Barrera$^4$, Ruochen Lu$^4$, and Gaurav Bahl$^{1}$ \\
   \footnotesize{$^1$ Department of Mechanical Science $\&$ Engineering, University of Illinois at Urbana–Champaign, Urbana, IL 61801 USA,} \\
    \footnotesize{$^2$ Department of Electrical $\&$ Computer Engineering, University of Illinois at Urbana–Champaign, Urbana, IL 61801 USA,} \\
    \footnotesize{$^3$ Department of Physics, University of Illinois at Urbana–Champaign, Urbana, IL 61801 USA,} \\
    \footnotesize{$^4$ Department of Electrical $\&$ Computer Engineering, The University of Texas at Austin, Austin, TX 78712 USA} \\
    \footnotesize{*jieuny@illinois.edu}
}

\date{}

\usepackage{float}

\usepackage[pdftex,colorlinks=true,linkcolor=blue,citecolor=blue]{hyperref}

\begin{document}
\begin{bibunit}

\maketitle

\begin{abstract}

Spatiotemporal modulation approaches have been often employed as alternatives for producing optical non-reciprocity without magneto-optic materials.
Unidirectional inter-modal scattering, enabled by either acousto-optic or electro-optic (EO) modulation, is a promising method in this category as it can directly modify optical dispersions and even enables linear non-reciprocal photonic devices in the strong coupling limit.
While EO approaches are often preferred for their practicality, it is challenging to generate the large spatiotemporal momentum required for inter-modal phase matching without EO drive schemes involving multiple drive stimuli.
Here, we demonstrate highly selective non-reciprocal inter-modal EO scattering enabled by a single high-index radiofrequency (RF) traveling wave stimulus.
Our experimental demonstration is performed on a thin-film lithium niobate integrated photonics platform, in which we engineer a slow-wave radiofrequency (SWRF) transmission line with an effective RF index $>9$ that natively generates the required RF momentum while simultaneously maintaining strong RF-optical mode overlap. 
By additionally engineering the interaction length, we achieve a directional $\sim 20$ dB non-reciprocal scattering contrast.
The SWRF architecture provides a scalable route to magnetic-free non-reciprocity and establishes momentum-engineered RF waves as a powerful tool for next-generation, fully integrated non-reciprocal photonic systems.

\end{abstract}

\vspace{12pt}

\begin{figure}[htp]
    \begin{adjustwidth*}{-1in}{-1in}
    \hsize=\linewidth
    \includegraphics[width=1.1\columnwidth]{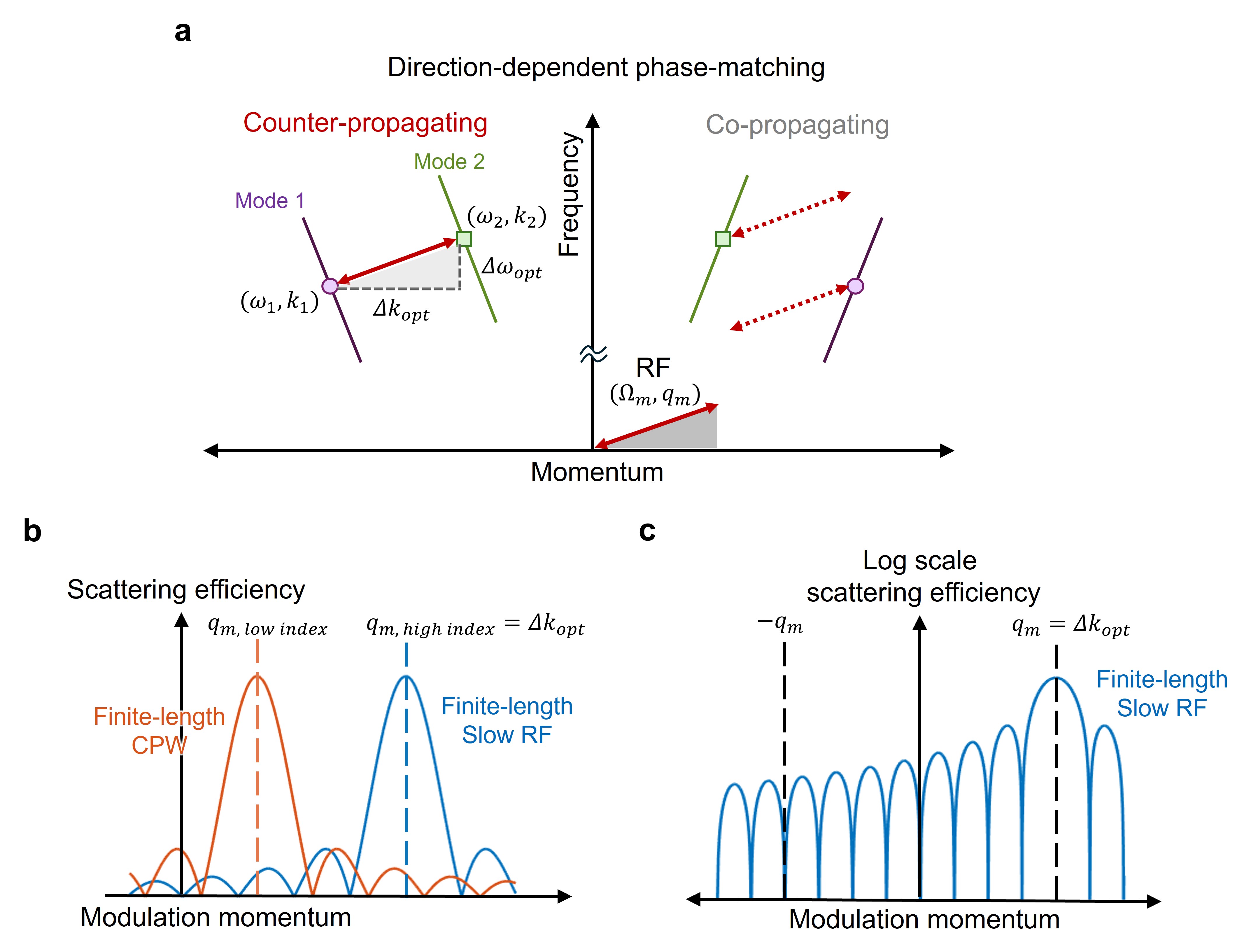}
    \centering
    \caption{\textbf{Principle of non-reciprocal inter-modal scattering by slow RF waves.} 
    \textbf{(a)} Diagram for direction-dependent phase matching of inter-modal scattering. Two optical modes can be coupled when their frequency difference $ \Delta\omega_{\text{opt}}$ and momentum difference $ \Delta k_{\text{opt}}$ match the frequency $\Omega_\textrm{m}$ and momentum $q_\textrm{m}$ of the RF wave, respectively. 
    In the specific configuration shown here, the directionality of the RF wave is only phase-matched with optical modes in the counter-propagating direction. In the co-propagating case, the phase-matching condition is not satisfied, and the scattering is suppressed.
 \textbf{(b)} Spectral scattering efficiency at a fixed frequency as a function of modulation momentum for a finite-length low-index CPW (orange) and a finite-length slow RF wave (blue). 
    \textbf{(c)} Same as (b), but plotted on a logarithmic scale, highlighting that the interaction length of slow RF wave can be intentionally designed such that the scattering efficiency exhibits a null at $-q_\textrm{m}$, thereby maximizing non-reciprocity. 
    }
    \label{fig1}
    \end{adjustwidth*}
\end{figure}

Non-reciprocal components such as isolators and circulators are essential building blocks in photonic systems. However, their on-chip integration is challenging due to the unavailability of Faraday rotator (i.e. magneto-optic) materials~\cite{bennett_faraday_1965,wolfe_thinfilm_1985} in photonics foundries. While there have been some research successes in incorporating magneto-optic materials into photonic integrated circuits~\cite{shoji_magneto-optical_2008,tien_silicon_2011,bi_-chip_2011}, more broadly accessible and foundry-compatible approaches are still needed---such as spatiotemporal modulation~\cite{kittlaus_electrically_2021,sohn_electrically_2021,tian_magnetic-free_2021,herrmann_mirror_2022,yu_integrated_2023,gao_thin-film_2024,kim_integrated_2025,cheng_terahertz-bandwidth_2025}.
 In this context, unidirectional inter-modal coupling has emerged as a promising alternative approach for breaking optical reciprocity~\cite{hwang_all-fiber-optic_1997,yu_complete_2009,yu_optical_2009,lira_electrically_2012}. 
 In this mechanism, two optical modes on distinct dispersion branches have frequencies and momenta $(\omega_1, k_1)$ and $(\omega_2, k_2)$, which are separated by ($\Delta\omega_{\mathrm{opt}}$, $\Delta k_{\mathrm{opt}}$) (Fig.~\ref{fig1}a). A traveling-wave stimulus can couple the modes when its frequency ($\Omega_\mathrm{m}$) and momentum ($q_\mathrm{m}$) match $\Delta\omega_\mathrm{opt}$ and $\Delta k_\mathrm{opt}$, respectively, thereby satisfying the phase-matching condition. Since $\Delta k_{\mathrm{opt}}$ reverses its sign for the opposite optical propagation direction, a fixed stimulus satisfies the phase-matching condition and produces inter-modal coupling in only one optical direction.
 This approach can be implemented through spatio-temporal modulation in both waveguides and resonators, by means of either the acousto-optic~\cite{sohn_time-reversal_2018,kittlaus_non-reciprocal_2018} or electro-optic effect~\cite{lira_electrically_2012,orsel_electro-optic_2023,kim_integrated_2025}.
Based on this principle, several architectures have been employed to build non-reciprocal devices~\cite{sohn_time-reversal_2018,kittlaus_non-reciprocal_2018,kittlaus_electrically_2021,sohn_electrically_2021,tian_magnetic-free_2021,orsel_electro-optic_2023,kim_integrated_2025}.

A key requirement for inter-modal scattering is a modulation stimulus with the correct momentum to bridge the optical momentum gap $\Delta k_\mathrm{opt}$, which tends to be large between distinct mode families. Acoustic waves have low phase velocity and can naturally provide the required momentum at modest frequencies ~\cite{sohn_time-reversal_2018,kittlaus_non-reciprocal_2018,kittlaus_electrically_2021,sohn_electrically_2021,sarabalis_acousto-optic_2021,zhang_integrated-waveguide-based_2024}. However, acousto-optic implementations typically require free surfaces or released structures, which complicates photonic integration, and scaling to higher acoustic frequencies is limited by fabrication constraints on the RF–acoustic transducers. Therefore, direct spatio-temporal modulation via the electro-optic (EO) effect has come to be of interest, as EO devices behave as capacitive electrical loads, can be fully cladded, and are readily scalable to high RF frequencies.

A major challenge, however, is that the RF phase velocity tends to be high in most coplanar waveguides (CPWs), which leads to a very low momentum at modest GHz frequencies ($< 100$ GHz), making it difficult to satisfy the inter-modal phase-matching requirement. 
One solution to address this issue is to employ a multi-phase or multi-drive split-electrode system~\cite{sounas_angular-momentum-biased_2014, orsel_electro-optic_2023, kim_integrated_2025} to synthetically engineer a large effective momentum that is determined by the electrode pitch.
However, it is very challenging to scale this approach to longer modulators or to higher RF stimulus frequencies since small deviations from the required phasing at each electrode can degrade the synthesized RF wave. Such phasing variations can readily arise from manufacturing tolerance issues, leading to capacitance variations, undesirable reflections, standing waves, and other RF engineering concerns.
Consequently, scalable electro-optic non-reciprocal devices may greatly benefit from an EO modulation scheme that is capable of natively generating a propagating wave with large momentum using only a single RF stimulus. 
    
In this work, we address this challenge by exploring a ``slow-wave'' radiofrequency (SWRF) transmission line structure~\cite{jaeger_slow-wave_1992, spickermann_millimetre_1993,spickermann_experimental_1994,sakamoto_narrow_1995} whose RF index can be adjusted while, in principle, simultaneously maintaining a desired waveguide impedance. Slow-wave electrodes have been used in Mach-Zehnder electro-optic modulators along with slotted and T-rail electrodes~\cite{li_analysis_2004,jaehyuk_shin_novel_2005,rosa_microwave_2018,huang_advanced_2021,kharel_breaking_2021,chen_high_2022,li_compact_2023} in which the goal is to match the optical group velocity with the RF phase velocity for intra-modal scattering. However, inter-modal scattering poses a more challenging problem, as it requires a transmission line with a very high RF index (low phase velocity) that simultaneously provides strong mode overlap between the RF field and two distinct optical modes~\cite{agarwal_multimode_2013,sohn_time-reversal_2018}.
Through the approach presented in this work, we address these challenges and demonstrate highly selective non-reciprocal inter-modal electro-optic scattering enabled by a single-drive SWRF transmission line.

\vspace{12pt}

As illustrated in the dispersion diagram (Fig.~\ref{fig1}a), we consider an optical waveguide system where inter-modal scattering is phase-matched when the optical and RF waves are counter-propagating. We note that this conclusion is design dependent and that the phase-matching may also be satisfied in the co-propagating direction with a suitable dispersion design. When the interaction length of the optical and RF waves is finite, the strict momentum-matching condition is relaxed, resulting in a sinc-shaped momentum response that peaks at $q_\mathrm{m}$ (see Supplementary \S{S1}). Figure~\ref{fig1}b presents an illustrative comparison of this sinc-shaped scattering efficiency of a finite-interaction-length system driven by low-index (high phase velocity) and high-index (low phase velocity) RF waves at a fixed frequency $\Omega_\mathrm{m}$. 
As the RF stimulus momentum $q_\mathrm{m}$ and index $n_\mathrm{RF}$ are related by $q_\mathrm{m}=2 \pi n_\mathrm{RF} \Omega_\mathrm{m} / c$, we find that a higher index RF transmission line naturally produces a larger momentum for the same stimulus frequency $\Omega_m$.
Fig.~\ref{fig1}c explores the scattering efficiency of the high-index case on a logarithmic scale, where the counter-propagating and co-propagating cases, corresponding to $q_\mathrm{m} = \Delta k_\mathrm{opt}$ and $q_\mathrm{m} = -\Delta k_\mathrm{opt}$, respectively, are indicated. Here, we note that it is possible to intentionally engineer the sinc-shaped spectral response to simultaneously place a maximum at $\Delta k_\mathrm{opt}$ and a null at $-\Delta k_\mathrm{opt}$ by choosing an appropriate interaction length $\mathcal{L} = p\pi/q_\mathrm{m}$ for any positive integer $p$, which maximizes the non-reciprocal scattering contrast (see Supplementary \S{S2}).

\begin{figure}[htp]
    \begin{adjustwidth*}{-1in}{-1in}
    \hsize=\linewidth 
    \includegraphics[width=1.2\columnwidth]{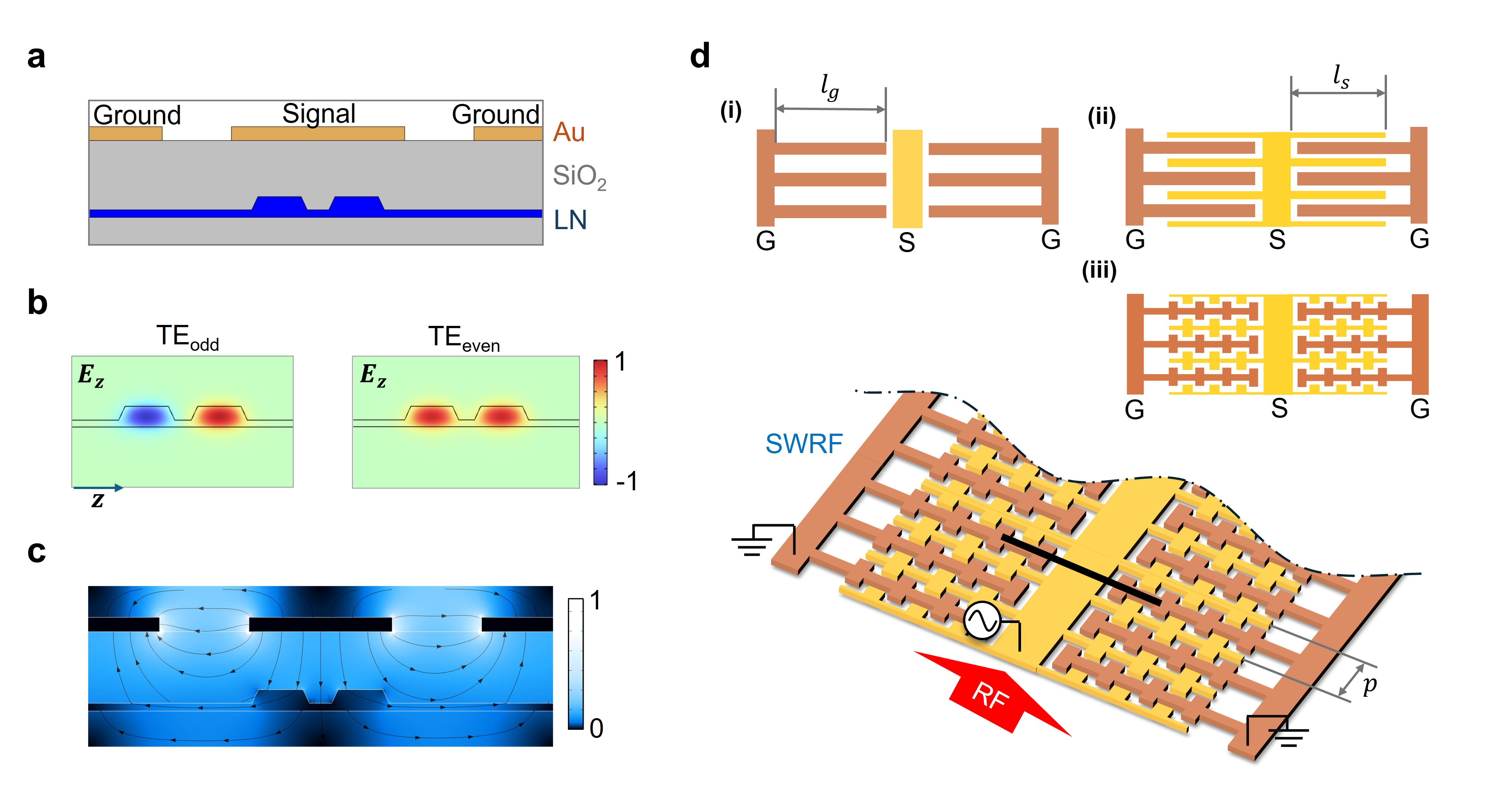}
    \centering
    \caption{
    \textbf{Design and implementation of our experimental device.} 
    \textbf{(a)} Cross-section schematic of the double optical waveguide system and the electro-optic modulation electrodes.
    \textbf{(b)} Simulated electric-field profiles ($E_z$) of the TE\textsubscript{odd} and TE\textsubscript{even} optical modes supported in the waveguide. 
    \textbf{(c)} Simulated radiofrequency field distribution from the electrodes. 
    \textbf{(d)} Schematic of the SWRF electrode. The insets (i)–(iii) illustrate the intermediate design steps of the SWRF electrode (bottom). The structure has periodicity of $p=10$ \um. The ground lines (G) have lateral slots of width $l_\textrm{g}=170$ \um. The signal lines (S) have lateral extensions of width  $l_\textrm{s}=36$ \um. The black solid line indicates the location of the cross-section whose RF field distribution is shown in Fig.~\ref{fig2}c.
    }
     \label{fig2}
    \end{adjustwidth*}
\end{figure}

Our experimental demonstration was conducted in a 600 nm X-cut thin film lithium niobate on insulator (LNOI) integrated photonics platform at 1550 nm (fabrication details are provided in Methods). 
We designed a double optical waveguide system (Fig.~\ref{fig2}a) that supports fundamental TE modes near 1550 nm with odd and even symmetry (Fig.~\ref{fig2}b) propagating along the crystal $\hat{y}$ axis of the lithium niobate.
This dual waveguide approach enables more fabrication-tolerant control over the momentum gap between the optical modes. The buried oxide was 2 \um thick and we deposited a 2 \um cladding oxide above the optical structures. Gold electrodes in a ground-signal-ground transmission line configuration were fabricated above the cladding to produce a mirror symmetric RF electric field for modulation (Fig.~\ref{fig2}c). 
With this choice of propagation direction and crystal orientation, the mirror symmetric RF modulation field maximally utilizes the largest Pockels coefficient of lithium niobate ($r_{33} \approx 30$ pm/V) to couple the $\hat{y}$ propagating TE optical modes \cite{kim_integrated_2025}. The transverse geometry of the electrodes (e.g. gap between the ground and the signal and the thickness of the cladding oxide) was designed to maximize the RF–optical overlap integral while minimizing optical loss induced by the metal electrodes.

The most important component in our experiment is the SWRF transmission line that respects the mirror symmetry with a strong mode overlap.
The transmission line inductance $L$ and capacitance $C$ per unit length provide two degrees of freedom to design both the modulation momentum $q_\textrm{m}$ and the characteristic impedance $Z_0$ according to $q_\textrm{m} = 2\pi\Omega_\textrm{m} \sqrt{LC}$ and $Z_0 = \sqrt{L/C}$.
Our final SWRF transmission line design is illustrated in Fig.~{\ref{fig2}}d, with the propagation direction indicated by the red arrow. The structure is composed of a periodic unit cell (with period $p$=10{$~\um$}), which features minimally interrupted straight sections that produce a local RF field (as visualized in Fig.~{\ref{fig2}}c) to match a quasi-TEM mode of a conventional CPW transmission line. This region is indicated with the black cross-section line in Fig.~{\ref{fig2}}d.
We manually derived the SWRF structure starting from a standard CPW line by increasing the inductance per unit length $L$ with the introduction of lateral slots of length $l_\mathrm{g}$ \cite{huang_advanced_2021} on the ground lines (inset (i) in Fig.~\ref{fig2}d), and increasing $C$ by introducing lateral extensions on the signal line with length $l_\mathrm{s}$ (inset (ii) in Fig.~\ref{fig2}d).  
At this stage, the resulting GSG transmission line appears as an interdigitated structure. Finally, the signal and ground extensions were patterned with alternating wide and narrow sections, rather than a uniform width, to increase the capacitance without further lengthening $l_\mathrm{s}$ (inset~(iii) in Fig.~{\ref{fig2}d}). 
The final geometry (details in Supplementary~\S{S4}) was optimized using full-wave simulations in Ansys HFSS and COMSOL Multiphysics, yielding an RF effective index of 9.8 and a characteristic impedance of $Z_0 = 42~\Omega$ over the frequency range $\Omega_\mathrm{m} = 16$--$20~\mathrm{GHz}$. 
This frequency range was chosen based on our experimental constraints on RF drive power, frequency, and measurement bandwidth, but the same design principles are readily applicable at higher RF frequencies.  
While the impedance can be further tuned by increasing $l_\mathrm{g}$, the present design was selected to remain compatible with the footprint of the optical structure. With the estimated RF index in this frequency range, the SWRF electrode should provide a modulation momentum peak near $q_{m,\mathrm{SWRF}} = 3300$--$4100~\mathrm{rad/m}$. 
 
\begin{figure}[htp] 
    \begin{adjustwidth*}{-1in}{-1in}
    \hsize=\linewidth
    \includegraphics[width=1.1\columnwidth]{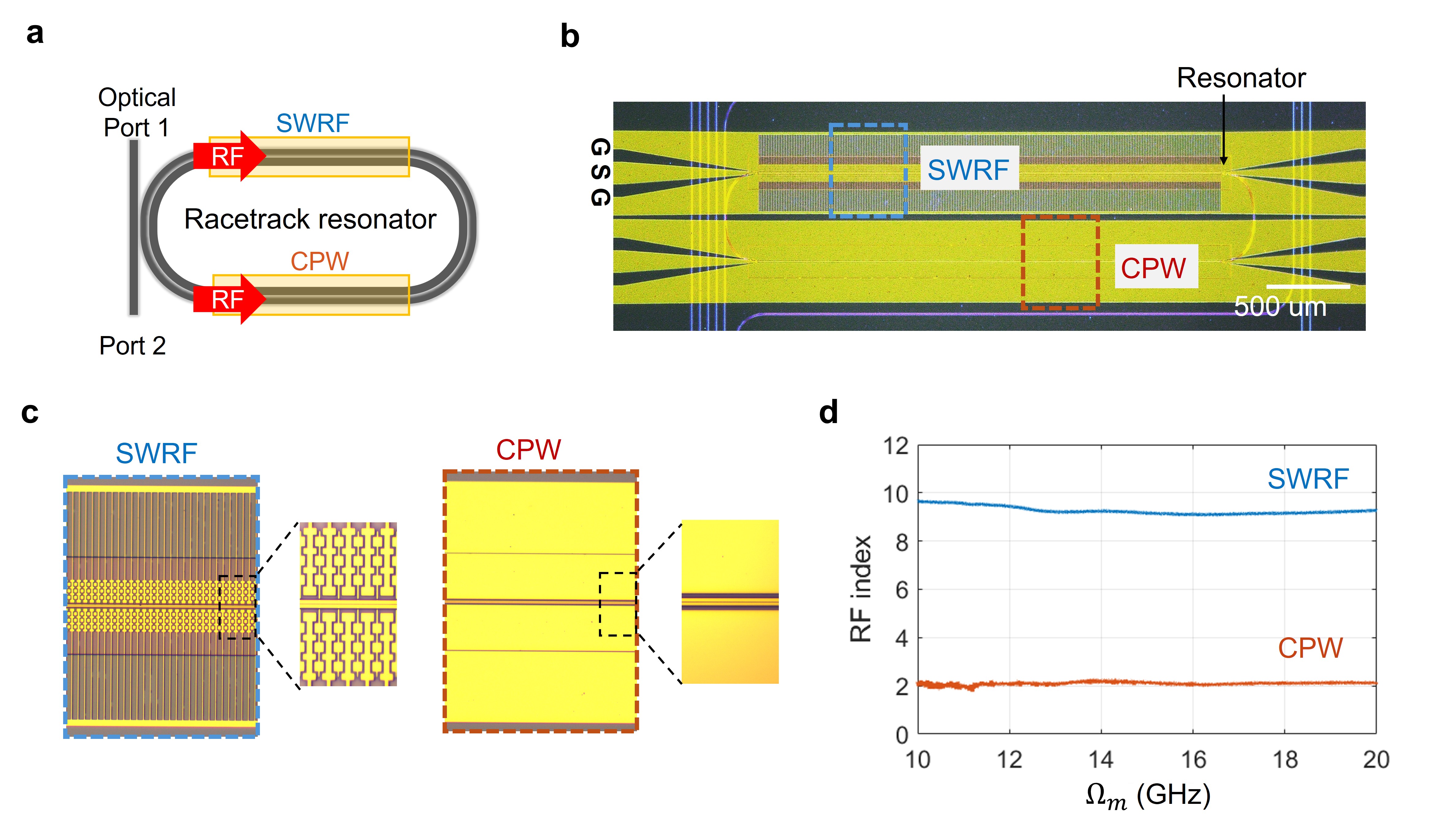}
    \centering
    \caption{
        \textbf{Images of the test device and RF characterization results.}
    \textbf{(a)} Schematic for the device consisting of a racetrack resonator and two modulators (SWRF and CPW electrodes) on two straight sections of the resonator. The optical signal enters through Port 1 and exits through Port 2, while the RF signals are launched from the ends of the electrodes. 
    \textbf{(b)} Optical microscope image of the fabricated device. Ground-Signal-Ground (GSG) pads are implemented at the ends of the electrodes via adiabatic tapers to interface with the RF input and output signals. 
    \textbf{(c)} Zoomed-in microscope images of the fabricated electrodes. The left panel shows the SWRF electrode with a periodic interdigitated structure, while the right panel shows the CPW. 
    \textbf{(d)} Extracted RF effective index as a function of RF frequency $ \Omega_m $, showing that the SWRF waveguide achieves a significantly higher index (above 9) compared to the CPW (around 2) over the 10--20~GHz range.
    }

    \label{fig3}
    \end{adjustwidth*}
\end{figure}

In order to directly compare the scattering efficiency of SWRF and CPW electrodes, we wrapped the dual-waveguide system into a double-racetrack resonator that supports the required TE{\textsubscript{even}} and TE{\textsubscript{odd}} modes, and fabricated both SWRF and CPW electrodes on the identical straight sections (schematic in Fig.~{\ref{fig3}}a and image in Fig.~{\ref{fig3}}b). A bus optical waveguide interfaces light into the resonator from edge couplers. The free spectral range (FSR) of the resonator was designed to be $\sim$18~GHz so that we can select and scatter non-adjacent TE{\textsubscript{even}} and TE{\textsubscript{odd}} modes separated by nearly one FSR (as later shown in the experiment). This enables inter-modal modulation at $\Omega_\mathrm{m}$ within the desired 16--20~GHz range.
The dispersions of the TE$_\text{even}$ and TE$_\text{odd}$ modes were then engineered by adjusting waveguide geometry such that the inter-modal momentum separation $\Delta k_\mathrm{opt}$ substantially overlaps with the achievable modulation momentum range ($q_{m,\mathrm{SWRF}}$ described above), while remaining compatible with design and fabrication constraints. The final geometry is detailed in Supplementary Fig.~S3. 
For an optical frequency separation range of $\Delta\omega_\mathrm{opt}=$ 16--20 GHz, the optimized dual-waveguide geometry exhibits $\Delta k_\mathrm{opt}$ spanning $3190$--$4190~\mathrm{rad/m}$ between the TE$_\text{even}$ and TE$_\text{odd}$ mode branches. 
For reference, the intra-modal momentum separation between adjacent resonant modes within each mode family is $758$--$951~\mathrm{rad/m}$, across the entire laser tuning range of 1520--1570~nm.
    
On the straight sections of the racetrack resonator, the SWRF and CPW modulator lengths were set to $\mathcal{L} = 2.85~\mathrm{mm}$. This value was determined (see Supplementary \S{S2}) by considering the simultaneous positioning of the peak and nulls in the momentum spectrum (described previously in Fig.~{\ref{fig1}}c) to aim for $>20$ dB scattering contrast over the targeted range of $q_{m,\mathrm{SWRF}}$. We note that, for a fixed value of $q_{m,\mathrm{SWRF}}$, multiple interaction lengths $\mathcal{L}$ can satisfy the nulling condition. Among these solutions, we selected the longest interaction length that could fit within the straight section of the resonator, which is limited by the FSR constraint.

The RF drive signals were supplied via a standard ground-signal-ground (GSG) configuration, with a 50 ohm load impedance applied on the opposite GSG pads to arrange for a traveling wave excitation. The GSG pads interface with the corresponding transmission lines via an adiabatically tapered section (Fig.~{\ref{fig3}b}) to mitigate undesirable reflections.
The relative direction of optical and RF signals can be readily switched by reversing the RF input and load, or by reversing the direction of input to the bus optical waveguide. 
We experimentally measured the index of the transmission lines using S-parameter analysis with a vector network analyzer. We used the general line-line method (see Supplementary \S{S5}) to extract the propagation constants by comparing the wave--cascade matrices of transmission lines of different lengths over a broad frequency range. 
The RF index thus measured over the 10--20~GHz range is shown in Fig.~\ref{fig3}d, confirming that the effective index of our CPW is $\sim$2, while the index of the SWRF electrode reaches $\sim$9.2 across the measured range, which is slightly lower than our simulated value.

\begin{figure}[htp]
    \begin{adjustwidth*}{-1in}{-1in}
    \hsize=\linewidth  
    \includegraphics[width=\columnwidth]{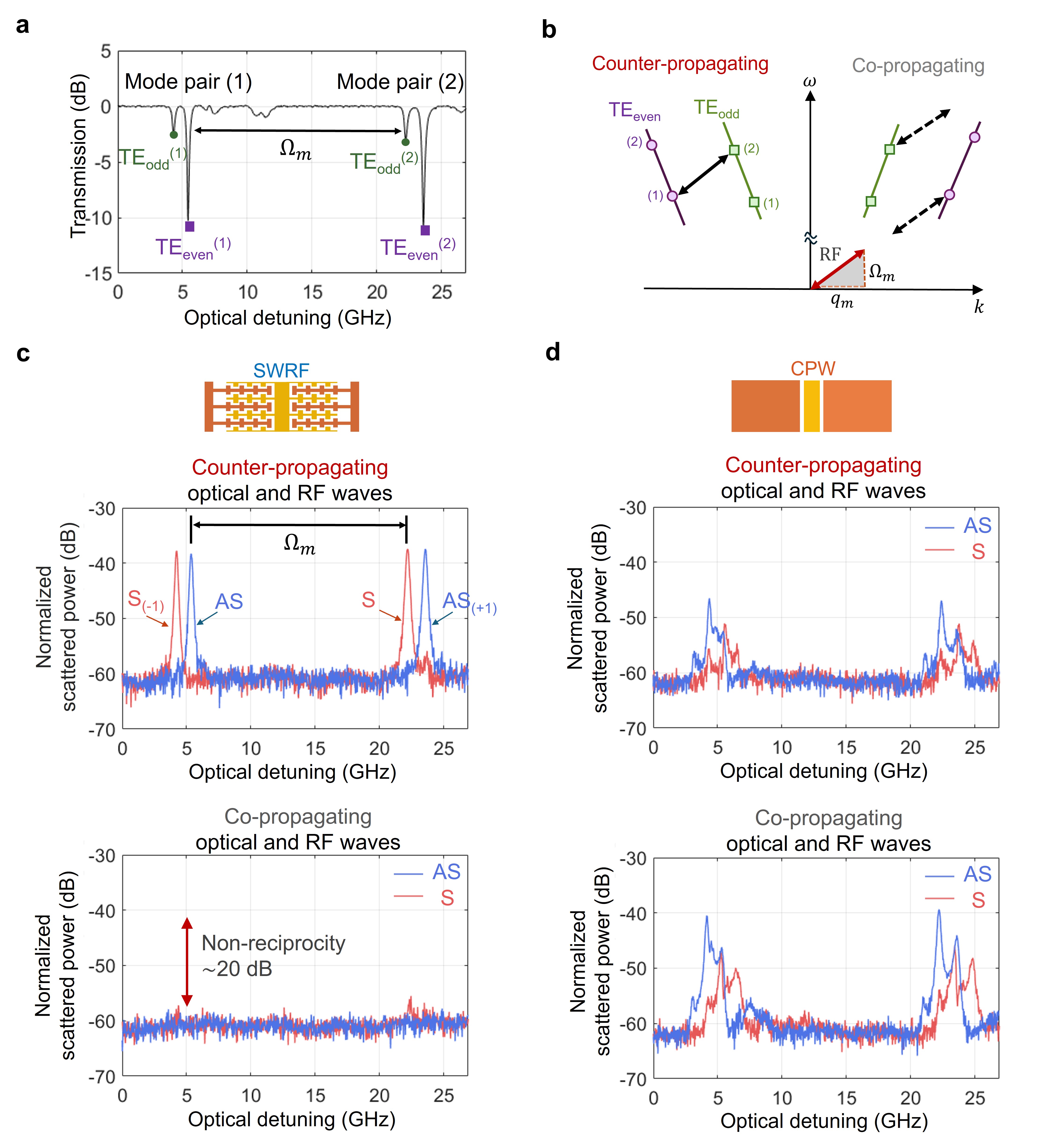}
    \centering
    \caption{\textbf{Experimental characterization of non-reciprocal scattering.} 
    \textbf{(a)} Measured optical transmission spectrum of the racetrack resonator, showing two mode pairs (Mode pair (1) and Mode pair (2)) separated by one FSR. The inner mode pair TE\textsubscript{even}\textsuperscript{(1)} and TE\textsubscript{odd}\textsuperscript{(2)} are separated by $16.8~\mathrm{GHz}$, corresponding to the RF modulation frequency $\Omega_\textrm{m}$ used in this study. 
    \textbf{(b)} The inferred positioning of the optical modes is indicated in this diagram, along with inter-modal scatterings enabled by the RF stimulus.
    \textbf{(c)} Measured and normalized scattered optical power under SWRF modulation. When the optical and RF signals are counter-propagating (top), strong non-reciprocal inter-modal scattering is observed, with well-resolved anti-Stokes (AS) and Stokes (S) sidebands. AS$_{(+1)}$ is the anti-Stokes scattering to the next FSR, and S$_{(-1)}$ is the Stokes scattering to the previous FSR. When the optical and RF signals are co-propagating (bottom), scattered power is suppressed to the noise floor, demonstrating $\sim 20~\mathrm{dB}$ non-reciprocity. 
    \textbf{(d)} Measured and normalized scattered power using the CPW modulator. In both counter- and co-propagating cases, we observe a cluttered spectrum with indications of mixture of inter-modal and intra-modal scattering.
    }
    \label{fig4}
    \end{adjustwidth*}
\end{figure}

We characterized the non-reciprocal scattering in our device by measuring the generated optical sidebands under RF modulation. We used a heterodyne detection setup (described in Methods) to measure the carrier, Stokes (down-converted), and anti-Stokes (up-converted) signals independently, while sweeping the laser input frequency with detuning over a 26 GHz span around the modes of interest.
Fig.~{\ref{fig4}}a shows a carrier transmission spectrum near 1547 nm through the optical bus waveguide and presents the set of optical modes used for subsequent experiments. The TE$_\mathrm{even}$ and TE$_\mathrm{odd}$ modes have similar intrinsic quality factor of $\approx 6\times 10^5$ and are identified by their relative coupling to the bus waveguide. In our double-racetrack resonator, the TE$_\mathrm{even}$ modes couple more strongly to the waveguide, resulting in a greater resonance dip. 
Fig.~{\ref{fig4}}b illustrates the organization of these modes in frequency-momentum space, as inferred from the transmission spectrum and our design of the dispersion. 

To measure the inter-modal scattering efficiency, we selected the inner mode pair labeled TE$_\mathrm{even}^{(1)}$ and TE$_\mathrm{odd}^{(2)}$, which has a frequency separation of $16.8~\mathrm{GHz}$. The RF drive frequency was therefore fixed at $\Omega_\mathrm{m}=16.8~\mathrm{GHz}$. During the modulation, the anti-Stokes and Stokes sideband powers were measured as a function of the input laser frequency.
At this drive frequency, the SWRF transmission line supplies a modulation momentum of 3240 rad/m, which is slightly below the targeted range of $q_{m,\mathrm{SWRF}}$. Despite this discrepancy, efficient inter-modal scattering in the phase-matched direction is still achievable for the designed optical dispersion, with a theoretically predicted non-reciprocal contrast that remains larger than $18.8~\mathrm{dB}$ in the sideband powers across the wavelength range of 1520--1570~nm (see Supplementary Fig. S6a).

Experimental results presented in Fig.~\ref{fig4}c show scattered sideband powers for the co-propagating and counter-propagating cases (indicated by black arrows in Fig.~{\ref{fig4}}b) using the SWRF transmission line.
When the slow-wave RF drive is applied in the counter-propagating direction relative to the optical input (upper plot in Fig.~\ref{fig4}c), the inter-modal phase-matching condition is satisfied, which leads to clearly distinguished anti-Stokes ($AS$) and Stokes ($S$) signal peaks. Specifically, a prominent anti-Stokes offset signal appears when TE$_\mathrm{even}^{(1)}$ is excited, indicating large coupling to TE$_\mathrm{odd}^{(2)}$ through the electro-optic modulation. This inter-modal coupling is further confirmed by the presence of Stokes offset signal when the laser input excites the TE$_\mathrm{odd}^{(2)}$ resonance. 
In contrast, when the slow-wave RF drive is applied in the co-propagating phase-mismatched direction relative to the optical signal (lower plot in Fig.~\ref{fig4}c), the scattered power is below our measurement noise floor, implying a scattering non-reciprocity of at least $20~\mathrm{dB}$ between the co- and counter-propagating cases. We note the existence of two additional scattering peaks in the phase-matched (counter-propagating) case. The Stokes signal at TE\textsubscript{odd}\textsuperscript{(1)} (labeled $S_{-1}$) appears due to scattering into the TE\textsubscript{even} mode that is one FSR lower in frequency. Similarly, the anti-Stokes signal at TE\textsubscript{even}\textsuperscript{(2)} (labeled $AS_{+1}$) appears due to scattering into the TE\textsubscript{odd} mode that is one FSR higher in frequency.

To compare against a lower-index RF stimulus, we repeated the measurements with the same drive frequency, $\Omega_\mathrm{m}=16.8~\mathrm{GHz}$, now applied to the CPW transmission line. In both the counter-propagating (upper plot in Fig.{~\ref{fig4}}d) and co-propagating (lower plot in Fig.~{\ref{fig4}}d) configurations, the measured response consists of complicated overlapping spectral features that are not cleanly resolved. In the ideal limit of purely intra-modal scattering, the Stokes and anti-Stokes sidebands would completely overlap with a spacing determined by the resonator FSR since the availability of optical modes for intra-modal scattering is symmetric and leads to symmetric up and down conversion. 
The partially overlapping spectral features here in Fig.~{\ref{fig4}}d therefore indicate the simultaneous presence of intra- and inter-modal processes under CPW modulation. By contrast, the distinctly separated sidebands observed under SWRF drive (Fig.~{\ref{fig4}}c) are characteristic of highly selective inter-modal scattering. 

In order to better understand these spectral features, we consider the sinc-shaped momentum spectrum of our finite length CPW waveguide. We calculate that the momentum spectrum peaks at $q_{\mathrm{m,CPW}} = 704~\mathrm{rad/m}$ which is much smaller than the inter-modal momentum separation $\Delta k_\mathrm{opt}$ (see Supplementary Fig.~S6b).

Consequently, the two relevant optical momentum spacings $\pm\Delta k_\mathrm{opt}$ lie on higher-order lobes of the sinc spectrum, where the scattering efficiency has lower asymmetry.
At the same time, because $q_{\mathrm{m,CPW}}$ is close to the intra-modal momentum separation, the CPW drive efficiently excites intra-modal scattering. The coexistence of these two mechanisms therefore leads to various interference effects and a cluttered scattering spectrum.
In conclusion, these results demonstrate that a slow-RF stimulus providing sufficiently large modulation momentum is essential for selectively exciting inter-modal scattering and achieving strong non-reciprocity.

\begin{figure}[H]
    \begin{adjustwidth*}{-1in}{-1in}
    \hsize=\linewidth 
    \includegraphics[width=1.3\columnwidth]{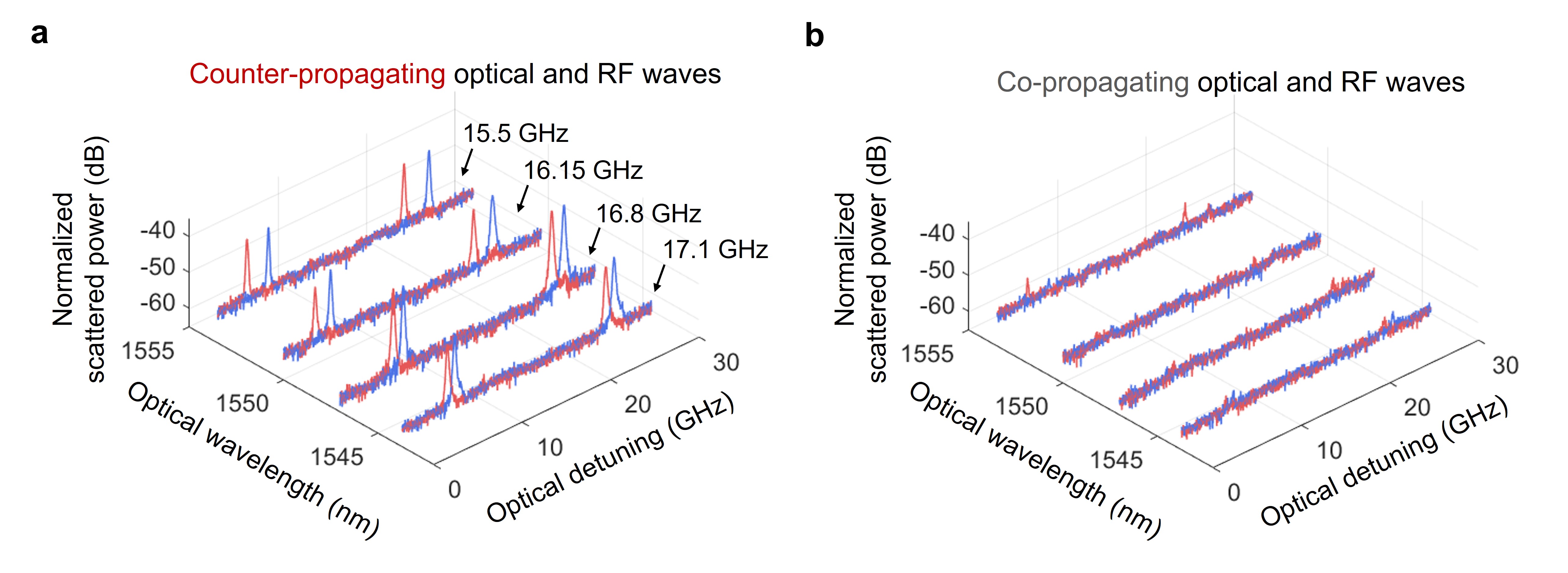}
    \centering
    \caption{ \textbf{Experimental characterization of non-reciprocal scattering via the SWRF waveguide at different optical wavelengths.} \textbf{(a)} Measured Stokes (red) and anti-Stokes (blue) sidebands when optical and RF waves are counter-propagating.
    \textbf{(b)} Measured Stokes (red) and anti-Stokes (blue) sidebands when optical and RF waves are co-propagating. Scattered power remains significantly suppressed, confirming robust non-reciprocal behavior across all tested wavelengths.
    }
    \label{fig5}
    \end{adjustwidth*}
\end{figure}

Finally, we also tested the SWRF modulator with optical mode sets at different wavelengths in the telecom band and confirmed the clean non-reciprocal behavior (Fig.~\ref{fig5}). We note that, due to dispersion, the frequency separation between the optical modes does vary slightly with wavelength, so the modulation frequency was accordingly adjusted from 15.5~GHz to 17.1~GHz. 
The appearance of clear anti-Stokes and Stokes sidebands in the counter-propagating case confirms strong and highly selective inter-modal scattering, in contrast to reduced scattering in the co-propagating case. The enhanced phase matching provided by the SWRF structure remains effective despite the dispersion.

\section*{Discussion}

In this work we demonstrate a new approach to single input electro-optic non-reciprocal intermodal scattering enabled via a slow wave radiofrequency (SWRF) waveguide design.
We show that a waveguide with large RF index can natively generate the modulation momentum required to bridge the inter-modal gap, enabling highly selective scattering without needing multi-stimulus configurations. 
In addition, we show that the finite interaction length provides a complementary degree of freedom that enables a simultaneous optimization of peak scattering for positive optical momenta and scattering nulls for negative momenta. 
This momentum-engineering framework can be extended to longer modulators and higher operating frequencies, as long as the electrode interdigitation period remains deep-subwavelength so that the transmission line behaves as an effectively uniform slow-wave medium. 
Looking beyond non-reciprocal photonics, momentum-engineered RF waves can enhance phase matching and spectral selectivity in integrated transducer designs, advancing microwave–optical transduction~\cite{andrews_bidirectional_2014,mirhosseini_superconducting_2020,shen_traveling-wave_2024}, a process central to quantum networking and coherent RF–photonics interfaces. This work opens new opportunities for RF-driven isolators, circulators, non-reciprocal mode converters, and other devices leveraging momentum-engineered RF waves in next-generation electro-optic systems.

\section*{Methods}
\label{sec:Method}

\textbf{Device fabrication --} The optical resonator and waveguides were first patterned on X-cut thin-film lithium niobate on a silicon dioxide layer on a silicon handle wafer using electron beam lithography. The developed pattern was then etched to a depth of 380~nm using argon-based inductively coupled plasma reactive ion etching (ICP-RIE). A 400~nm-thick gold electrode was subsequently fabricated on top of the 2 {\textmu}m-thick silicon dioxide cladding through electron beam lithography followed by a lift-off process. After the device is fully fabricated on the chip, the edge of the chip was cleaved to enable edge-coupling of light from a lensed fiber to an inverse tapered waveguide.

\vspace{12pt}

\noindent
\textbf{Heterodyne measurement system --} We built a heterodyne interferometer to detect optical sidebands generated from the input light with GHz-level frequency separation (see Supplementary \S{S6}). A tunable external cavity diode laser (New Focus TLB6700, 1520--1570~nm) was used as the optical source. For the reference arm, an acousto-optic frequency shifter (Brimrose AMF-100-8-1550-2FP) was used to generate a 100~MHz frequency-shifted signal. The resulting optical sidebands were measured using a high-speed photodetector (Thorlabs RXM38AF) that operates up to 38~GHz and a vector network analyzer (Rohde \& Schwarz ZNA67), which also supplied the RF drive for modulation up to 18~dBm.

\end{bibunit}

%
\section*{Acknowledgments}

This work was sponsored by the Defense Advanced Research Projects Agency (DARPA) under Cooperative Agreement D24AC00003, the US Office of Naval Research (ONR) Multi-University Research Initiative grant N00014-20-1-2325, the Army Research Office (ARO) grant W911NF-23-1-0219, and National Science Foundation (NSF) under CAREER Award No. 2339731. The views and conclusions contained herein are those of the authors and should not be interpreted as necessarily representing the official policies or endorsements, either expressed or implied, of DARPA, ONR, ARO, NSF or the US Government.

\newpage

\newpage

\renewcommand*{\citenumfont}[1]{S#1}
\renewcommand*{\bibnumfmt}[1]{[S#1]}
\newcommand{\beginsupplement}{%
        \setcounter{table}{0}
        \renewcommand{\thetable}{S\arabic{table}}%
        \setcounter{figure}{0}
        \renewcommand{\thefigure}{S\arabic{figure}}%
        \setcounter{equation}{0}
        \renewcommand{\theequation}{S\arabic{equation}}%
        \setcounter{section}{0}
        \renewcommand{\thesection}{S\arabic{section}}%
}

\beginsupplement

\begin{bibunit} 

\begin{center}

\Large{\textbf{Supplementary Information:} \\ {Non-reciprocal electrooptic intermodal scattering with momentum engineered RF waves}} \\
\vspace{12pt}
\vspace{12pt}
\large{
Jieun Yim$^{1*}$, Gwan In Kim$^2$, Violet Workman$^3$, Seho Kim$^2$, Omar A. Barrera$^4$, Ruochen Lu$^4$, and Gaurav Bahl$^{1}$} \\
\vspace{12pt}
    \footnotesize{$^1$ Department of Mechanical Science $\&$ Engineering, University of Illinois at Urbana–Champaign, Urbana, IL 61801 USA,} \\
    \footnotesize{$^2$ Department of Electrical $\&$ Computer Engineering, University of Illinois at Urbana–Champaign, Urbana, IL 61801 USA,} \\
    \footnotesize{$^3$ Department of Physics, University of Illinois at Urbana–Champaign, Urbana, IL 61801 USA,} \\
    \footnotesize{$^4$ Department of Electrical $\&$ Computer Engineering, The University of Texas at Austin, Austin, TX 78712 USA} \\
    \footnotesize{*jieuny@illinois.edu}
    
\end{center}

\section*{S1. Scattering efficiency}

Here, we present how the scattering efficiency is determined by the RF momentum spectrum. The spatial profile of the RF traveling wave that interacts with the optical field during the propagation in the $y$ direction (for example, crystal y-axis in the main text) over the interaction length $\mathcal{L}$ (i.e. a rectangular spatial window) can be expressed as
\begin{equation}
    f(y) = A e^{i q_\mathrm{m} y} \cdot \mathrm{rect}(y),  \quad \text{where   } \mathrm{rect}(y) =
    \begin{cases}
        1, & \text{if } 0 < y < \mathcal{L} \\
        0, & \text{otherwise}
    \end{cases}
\end{equation}
where $A$ is the RF signal amplitude and $q_\mathrm{m}$ is the momentum of RF wave. By applying Fourier transform to \(f(y)\), we obtain the momentum spectrum of the RF wave written as
\begin{equation}
\label{eq:Fq}
F(q)=A_{q}\mathcal{L} \cdot \textrm{sinc} \left( \frac{(q-q_\mathrm{m})\mathcal{L}}{2} \right).
\end{equation}
This sinc function is centered at \(q=q_\mathrm{m}\), which is determined by the RF index according to \(q_\mathrm{m}=\frac{2\pi n_{RF}\Omega_\textrm{m}}{c}\) (\(n_{RF}\): effective index), and has zeros at \(q_n=q_\mathrm{m} \pm \frac{2n\pi}{\mathcal{L}}\) ($n\not=0$). 

The inter-modal scattering efficiency is then defined as the square of the total electro-optic coupling rate $\beta$ ($\frac{1}{s}$) between the two optical modes. For a traveling-wave modulation, the scattering efficiency scales as
\begin{equation}
    |\beta|^2
        \;\propto\;
    |\frac{\omega_0\Delta n_\textrm{EO}}{n_g}\,
    \cdot F(q)|^2
\label{eqs3}
\end{equation}
where $\omega_0$ is the optical frequency, and $n_g$ is the optical group index. $\Delta n_\mathrm{EO}$ is the electro-optically induced refractive index change, which is calculated by the overlap integral between the optical modes and the RF field via the relation~{\cite{orsel_electro-optic_2023}}
    \begin{equation}
    \Delta n_{\mathrm{EO}}
    =
    \frac{
    -\dfrac{\varepsilon_0 n^5}{2}
    \displaystyle
    \int
    \left[
    \mathcal{E}^{\mathrm{even}}(\mathbf{r}_\perp)
    \right]^{H}
    \, \boldsymbol{r}_{\mathrm{EO}} \,
    \left[
    \mathcal{E}^{\mathrm{RF}}(\mathbf{r}_\perp)
    \right]
    \left[
    \mathcal{E}^{\mathrm{odd}}(\mathbf{r}_\perp)
    \right]
    \, d\mathbf{r}_\perp
    + \mathrm{c.c.}
    }{
    \displaystyle
    \int
    \left(
    \mathcal{E}^{\mathrm{even}}(\mathbf{r}_\perp)
    +
    \mathcal{E}^{\mathrm{odd}}(\mathbf{r}_\perp)
    \right)
    \cdot
    \left(
    \mathcal{D}^{\mathrm{even}}(\mathbf{r}_\perp)
    +
    \mathcal{D}^{\mathrm{odd}}(\mathbf{r}_\perp)
    \right)
    \, d\mathbf{r}_\perp
    }
    \end{equation}
Here, $n$ is the refractive index of the material, and $H$ denotes the Hermitian conjugate. The $\boldsymbol{r}_{\mathrm{EO}}$ represents the third-rank electro-optic tensor of the material. $\mathcal{E}^{\mathrm{even}}(\mathbf{r}_\perp)$, $\mathcal{E}^{\mathrm{odd}}(\mathbf{r}_\perp)$, and $\mathcal{E}^{\mathrm{RF}}(\mathbf{r}_\perp)$ denote the transverse electric fields of the TE$_{\mathrm{even}}$, TE$_{\mathrm{odd}}$ optical modes, and the RF wave, respectively. Similarly, $\mathcal{D}^{\mathrm{even}}(\mathbf{r}_\perp)$ and $\mathcal{D}^{\mathrm{odd}}(\mathbf{r}_\perp)$ are the corresponding electric displacement fields of the optical modes. The denominator therefore represents the total electromagnetic energy normalization of the interacting modes over the transverse cross-section. Consequently, according to Eq.~{\ref{eqs3}}, for a fixed RF drive and RF-optical mode overlap, the electro-optically induced inter-modal scattering at any momentum spacing is governed by the squared momentum spectrum function $|F(q)|^2$.

\section*{S2. Determination of modulator length}
 
\begin{figure}[H]
  \centering
  \makebox[\textwidth][c]{%
    \includegraphics[width=1.1\textwidth]{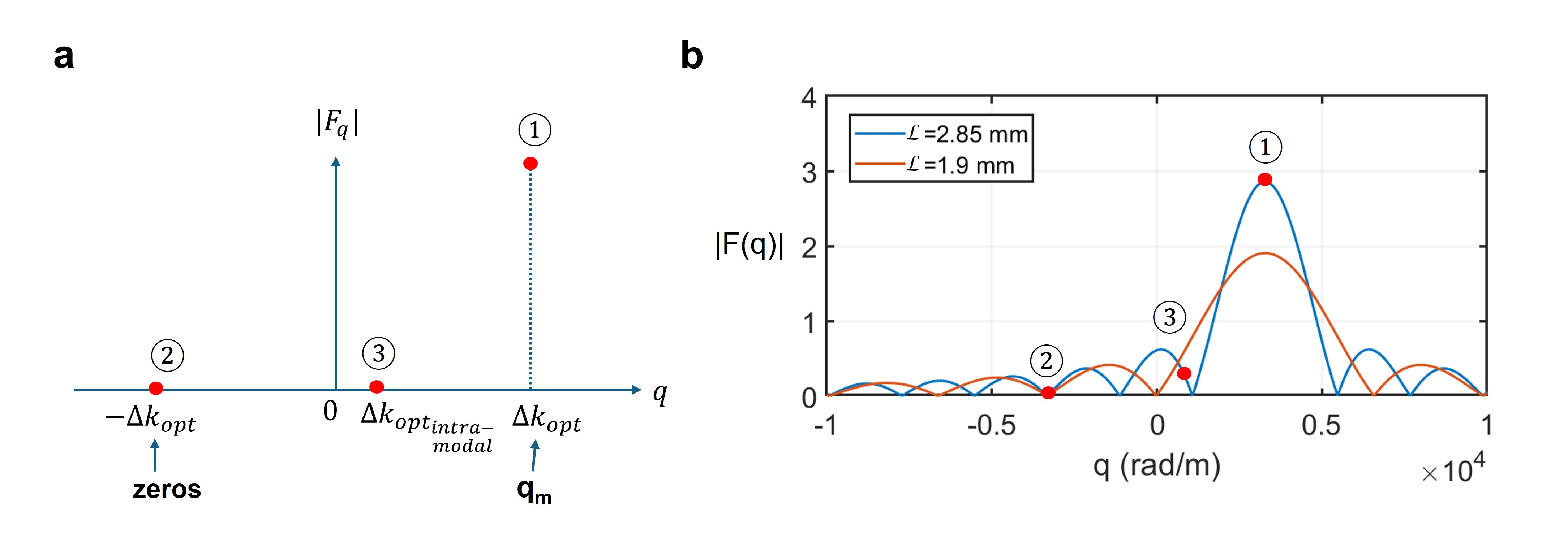}%
  }
\caption{\textbf{(a)}~Desired values that guide the momentum spectrum design.
         \textbf{(b)}~Example momentum spectra for two different modulator lengths $\mathcal{L}$.}
  \label{figS1}
\end{figure}

\begin{figure}[H]
  \centering
  \makebox[\textwidth][c]{%
    \includegraphics[width=0.6\textwidth]{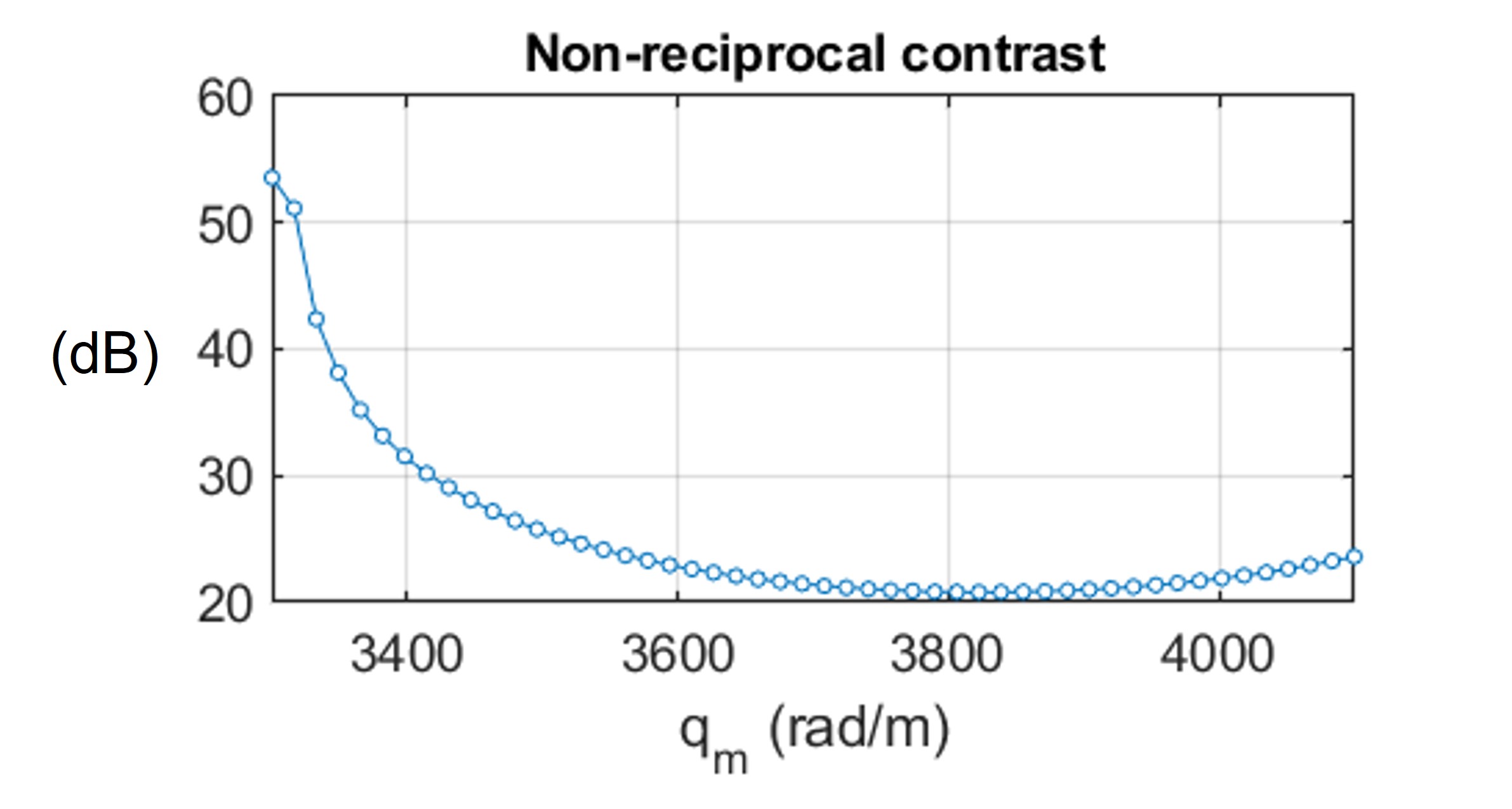}%
  }
\caption{Non-reciprocal contrast calculated from $\nicefrac{|F(q=\Delta k_\textrm{opt})|^2}{|F(q=-\Delta k_\textrm{opt})|^2}$ for modulator length $\mathcal{L}=2.85$ mm within a range of modulation momentum $q_m=3300 - 4100$ rad/m.}
  \label{figS1-2}
\end{figure}

In order to determine the ideal length $\mathcal{L}$ for the modulator, we consider the momentum spectrum $F(q)$. First, to maximize the inter-modal scattering, we should align \(q_\mathrm{m}\) with \(\Delta k_{opt}\), which is denoted as \circled{1} in Fig.~\ref{figS1}a. When the relative direction of optical and RF waves is reversed, the momentum response is determined by \(F(q=-\Delta k_{opt}\)). To null the reverse direction response for the maximum non-reciprocity, \(-\Delta k_{opt}\) should be located at a zero of the sinc function (\circled{2} in Fig.~\ref{figS1}a). Finally, to minimize intra-modal scattering, \(F(q=-\Delta k_{\mathrm{opt}_{\mathrm{intramodal}}})\) should also be located near a zero (\circled{3} in Fig.~\ref{figS1}). By substituting conditions \circled{1} and \circled{2} into Eq.~\eqref{eq:Fq}, we can derive the interaction length as 
\begin{equation}
\mathcal{L}=\frac{p\pi}{q_\mathrm{m}}
\end{equation}
where $\textrm{p}$ is a positive integer that can be determined arbitrarily as in Fig.~\ref{figS1}b. In our design, we use $q_\mathrm{m}=3300$ rad/m and $p=3$ which leads to $\mathcal{L}=2.85$ mm, by which we can also fulfill the requirement \circled{3}.
Since we are interested in the non-reciprocal contrast of the spectral scattering efficiency, which is expressed as the ratio $\nicefrac{|\beta(q=\Delta k_\mathrm{opt})|^2}{|\beta(q=-\Delta k_\mathrm{opt})|^2}$, 
we can instead compute the directly proportional ratio $\nicefrac{|F(q=\Delta k_\mathrm{opt})|^2}{|F(q=-\Delta k_\mathrm{opt})|^2}$ with $\mathcal{L}=2.85$ mm.
(see Fig.~\ref{figS1-2}). 
Since our experiments invoke a modulation frequency range of $\Omega_\textrm{m}=16-20$ GHz, the corresponding modulation momentum range is $q_m=3300 - 4100$ rad/m as described in the main text.
Fig.~{\ref{figS1-2}} shows that, for our chosen interaction length, the non-reciprocal contrast does vary with the modulation momentum $q_\mathrm{m}$, but remains robustly above $20~\mathrm{dB}$ across the entire range.

\section*{S3. Dual-waveguide design}
\begin{figure}[H]
  \centering
    \includegraphics[width=0.8\textwidth]{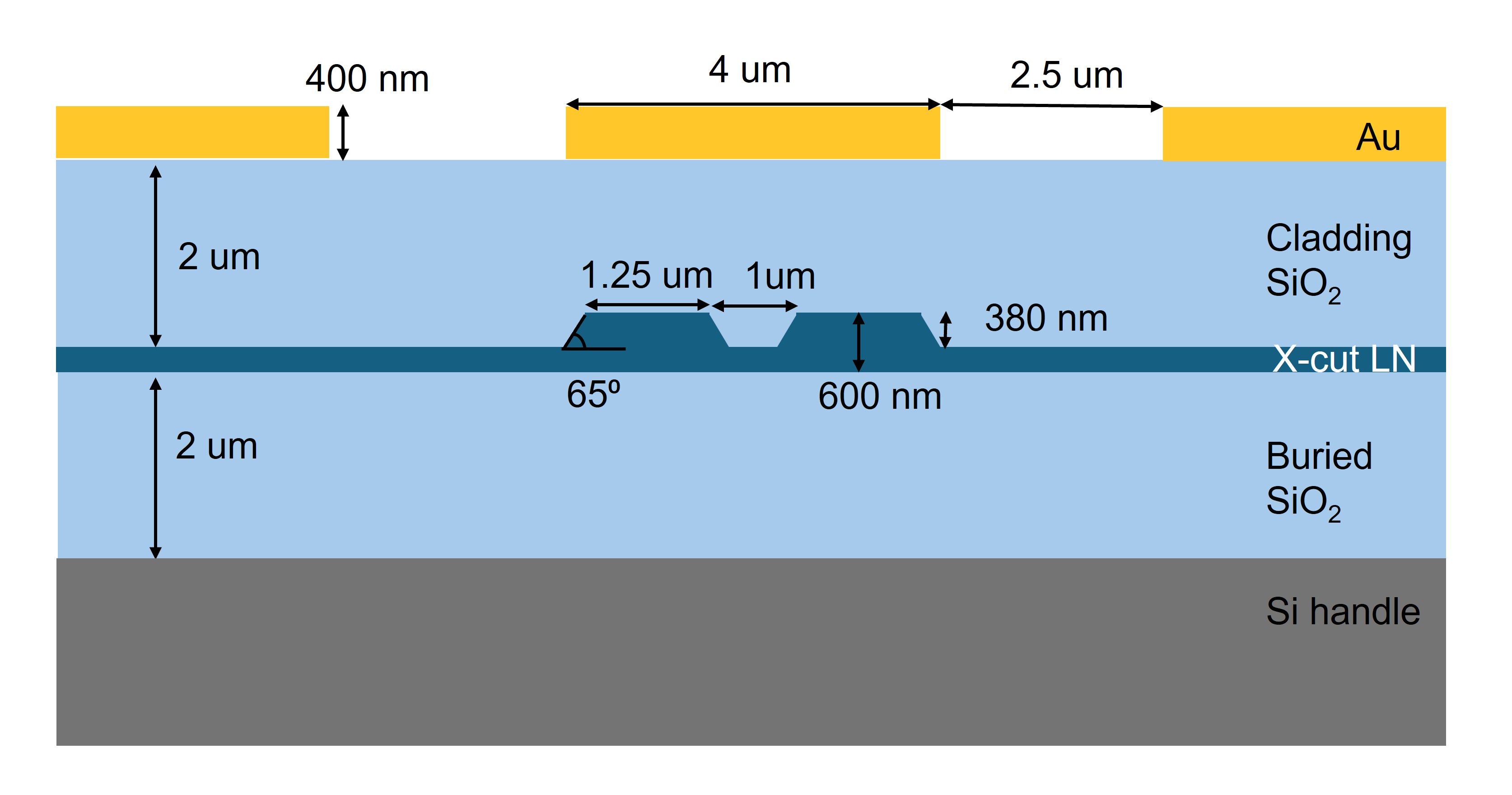}%
  \caption{Cross-sectional schematic of the double waveguide system and electrodes used in our experiments.}
  \label{figS2}
\end{figure}

Mode analysis was performed in COMSOL Multiphysics to design the dual-waveguide geometry. The optimized cross section of the dual optical waveguide is shown in Fig.~\ref{figS2}b. Several practical constraints were incorporated in the design. The optical waveguide separation must be sufficiently small to enable evanescent coupling that produces a resolvable splitting between the TE\textsubscript{even} and TE\textsubscript{odd} modes in the transmission spectrum, even in the presence of finite linewidths set by the loaded quality factors. At the same time, the optical waveguide width cannot be excessively large, as this would support higher-order modes. Conversely, the optical waveguide width cannot be too small, and the lithium niobate etch depth cannot be excessively deep, since increased optical mode interaction with etched sidewalls degrades the optical quality factor. Consequently, both design and fabrication constraints were jointly considered in selecting the final geometry.

\section*{S4. Dimensions of electrodes}
\begin{figure}[H]
  \centering
    \includegraphics[width=0.6\textwidth]{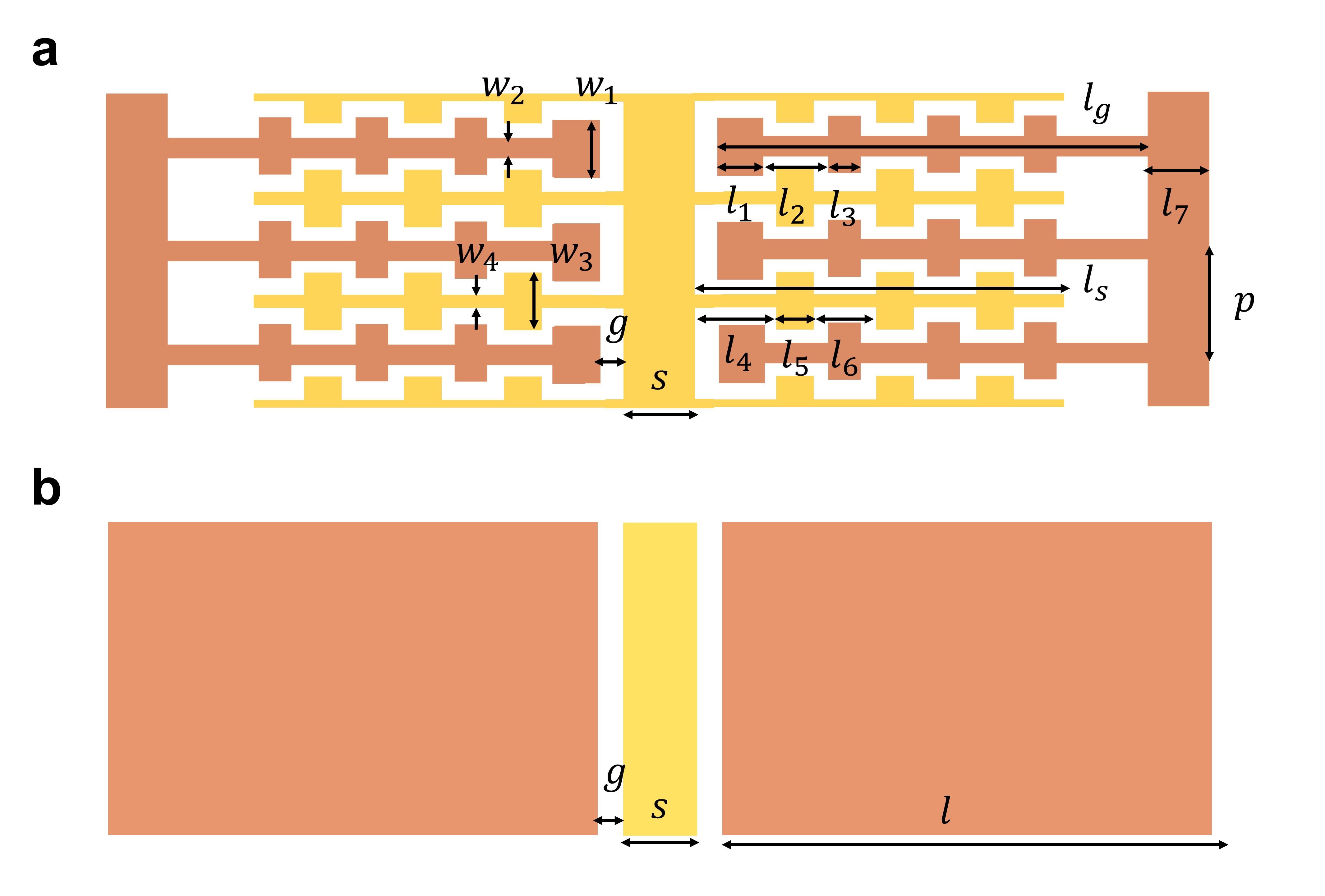}
  \caption{Dimensions of \textbf{(a)} SWRF \textbf{(b)} CPW electrodes. 
  The specific dimensions are listed: 
$s = 4~$\um, 
$g = 2.5~$\um, 
$p = 10~$\um, 
$l = 180~$\um,
$l_g = 170~$\um, 
$l_1 = 4.5~$\um, 
$l_2 = 7~$\um, 
$l_3 = 3~$\um, 
$l_4 = 8~$\um, 
$l_5 = 5~$\um, 
$l_6 = 5~$\um, 
$l_7 = 10~$\um,
$w_1 = 7~$\um, 
$w_2 = 3~$\um, 
$w_3 = 5~$\um, 
$w_4 = 1~$\um.
 } 
  \label{figS3}
\end{figure}

Through the design principle described in the main text, we performed 3D simulations and optimized the design of the electrode. The electrode thickness was set to 400 nm. The dimensions of the SWRF and CPW electrodes used in our device are shown in Fig.~\ref{figS3}.

\section*{S5. General line-line method}

To extract the effective RF index of the transmission lines, we employ the general line--line method~\cite{bao_general_2018}. Here, two transmission lines with identical cross-sectional geometry (SWRF or CPW) are fabricated, differing only in physical length. Under this condition, the propagation constant can be extracted directly from the length difference.
Let the two lines have lengths $\mathcal{L}_1$ and $\mathcal{L}_2$, with $\Delta \mathcal{L} = \mathcal{L}_2 - \mathcal{L}_1$. Their measured wave--cascade matrices are denoted $M_1$ and $M_2$, referenced to the same reference impedance (50~$\Omega$). These matrices can be expressed as
    \begin{align}
            M_1 &= X \, T(\mathcal{L}_1) \, Y, \\
            M_2 &= X \, T(\mathcal{L}_2) \, Y,
    \end{align}
where $X$ and $Y$ represent input and output networks (including RF probes, cables, and pads), and $T(\mathcal{L})$ is the wave--cascade matrix of the uniform transmission line section.

Forming the product as
    \begin{equation}
            M = M_2 M_1^{-1}
            = X \, T(\mathcal{L}_2) T(\mathcal{L}_1)^{-1} X^{-1},
    \end{equation}
eliminates the unknown network $Y$. Due to the trace invariance,
\begin{equation}
\mathrm{Tr}(M_2 M_1^{-1})=\mathrm{Tr}(T(\mathcal{L}_2)T(\mathcal{L}_1)^{-1}) .
\end{equation}

For a uniform transmission line of length $\mathcal{L}$ and complex propagation constant $\gamma$, the corresponding wave--cascade matrix has eigenvalues $e^{\pm \gamma \mathcal{L}}$, associated with the forward- and backward-propagating waves.
Consequently, for the length difference $\Delta \mathcal{L}$,
\begin{equation}
\mathrm{Tr}(M)
= e^{\gamma \Delta \mathcal{L}} + e^{-\gamma \Delta \mathcal{L}}
= 2 \cosh\!\left( \gamma \Delta \mathcal{L} \right).
\end{equation}
The complex propagation constant is thus extracted as
\begin{equation}
\gamma(\omega)
= \frac{1}{\Delta \mathcal{L}}
\cosh^{-1}\!\left( \frac{\mathrm{Tr}(M)}{2} \right),
\end{equation}
where the branch of the inverse hyperbolic cosine is chosen such that $\mathrm{Re}[\gamma] \ge 0$. The RF effective index is then obtained from the imaginary part of $\gamma$,
\begin{equation}
n_{\mathrm{RF}}(\omega)
= \frac{\mathrm{Im}[\gamma(\omega)]}{\omega / c}.
\end{equation}

\section*{S6. Heterodyne measurement and normalization}

\begin{figure}[H]
  \centering
  \makebox[\textwidth][c]{%
    \includegraphics[width=1.2\textwidth]{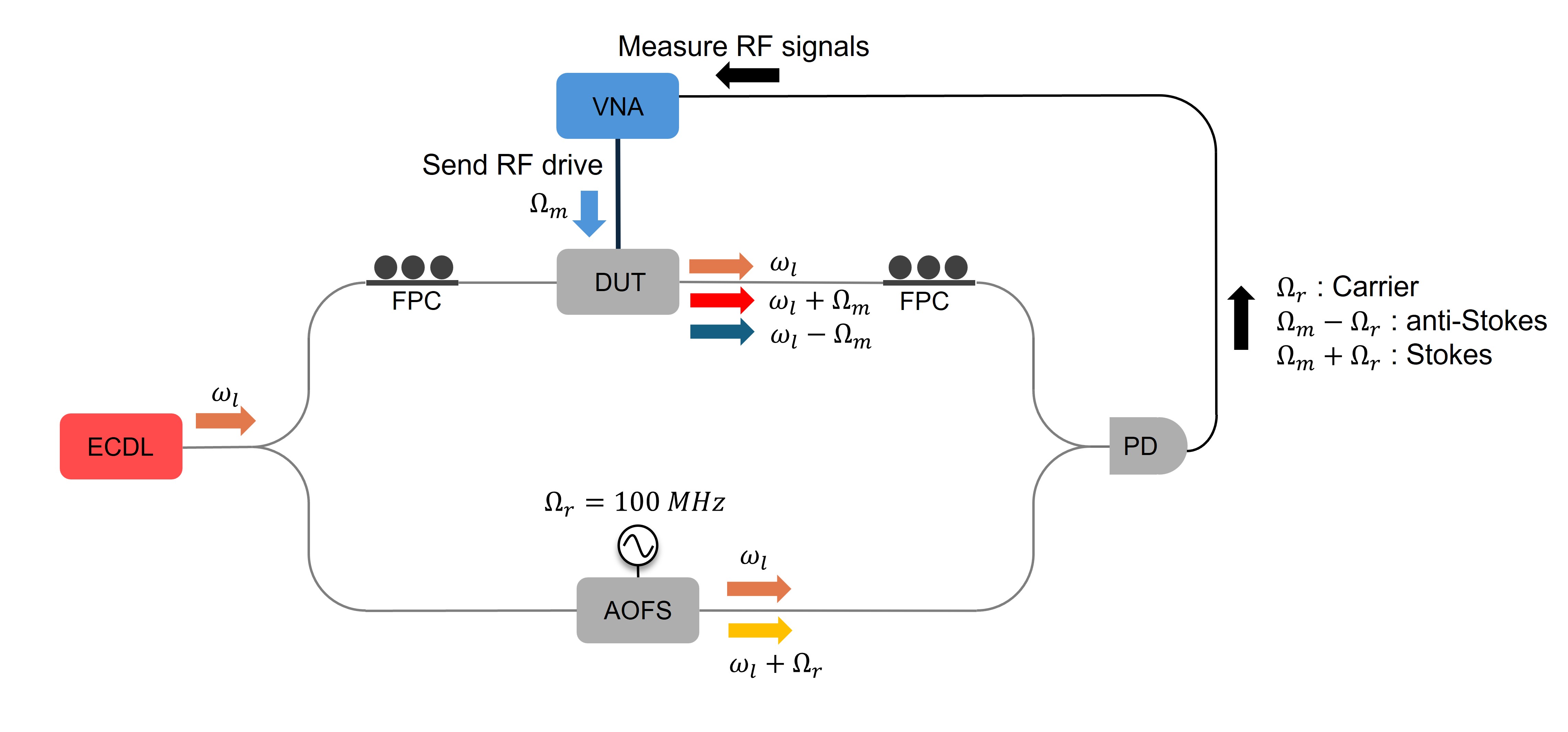}%
  }\caption{ Schematic of the heterodyne detection setup used to characterize anti-Stokes and Stokes sidebands. 
 ECDL: External cavity diode laser, DUT: device under test, VNA: Vector network analyzer, FPC: Fiber polarization controller, AOFS: Acousto-optic frequency shifter. PD: Photodetector. \(\omega_l\): Laser input frequency, \(\Omega_\textrm{r}\): reference frequency, \(\Omega_\textrm{m}\): RF modulation frequency. The direction of RF modulation can be reversed by interchanging the transmitting and receiving ports of the VNA. All the RF signals detected by PD were measured by the VNA.
}
  \label{figS4}
\end{figure}

As described in Methods in the main text, we build a heterodyne setup (Fig.~\ref{figS4}) to measure the scattered sideband powers relative to the input light, capable of measurements up to 26 GHz.

At the photodetector (PD), the detected optical field contains contributions from the laser output together with the anti-Stokes (AS), Stokes (S), and reference tones. Therefore, the optical field that arrives at the PD is described as
\begin{equation}
    E(t) =  S_{\text{out},0} e^{-i\omega_l t}
    + S_{\text{out},+1} e^{-i(\omega_l+\Omega_\textrm{m})t}
    + S_{\text{out},-1} e^{-i(\omega_l-\Omega_\textrm{m})t}
    + S_\textrm{r} e^{-i(\omega_l+\Omega_\textrm{r})t}
\end{equation}
Here, $S_{\text{out},0}$, $S_{\text{out},+1}$, and $S_{\text{out},-1}$ denote the amplitudes of the carrier, AS, and S optical fields, and their oscillating frequencies are $\omega_l$, $\omega_l+\Omega_\mathrm{m}$, and $\omega_l-\Omega_\mathrm{m}$, respectively. $S_\textrm{r}$ denotes the amplitude of the reference field oscillating at $\omega_l+\Omega_r$. These different frequency components and their pathways are illustrated in Fig.~{\ref{figS4}}.
Since the photocurrent generated by the photodetector is proportional to the incident optical power, $i(t) \propto |E(t)|^2$, interference between the reference field and the carrier transmission and first-order AS and S sidebands produces distinct RF beat-note
photocurrent components at the corresponding beat frequencies as
    \begin{equation}
        i_{\Omega_\mathrm{r}}(t)
                \propto 
        2\Re \{ S_{\text{out},0} S_\textrm{r}^{*} e^{i(\Omega_\textrm{r})t} \}
    \end{equation}
    \begin{equation}
        i_{\Omega_\textrm{m} - \Omega_\textrm{r}}(t)
            \propto 
        2 \Re \{ S_{\text{out},+1} S_\textrm{r}^{*} e^{-i(\Omega_\textrm{m}-\Omega_\textrm{r})t} \}
    \end{equation}
    \begin{equation}
    i_{\Omega_\textrm{m} + \Omega_\textrm{r}}(t)
    \propto 
    2 \Re \{ S_{\text{out},-1} S_\textrm{r}^{*} e^{-i(\Omega_\textrm{m}+\Omega_\textrm{r})t} \}
    \end{equation}
As a result, the carrier transmission, AS and S signal powers can be acquired at the vector network analyzer (assuming total scaling factor including PD gain to be $g_\text{pd}$) as
    \begin{equation}
        P_{\Omega_\textrm{r}} = g_{\text{pd}} |S_{\text{out},0}|^2 |S_\textrm{r}|^2
    \label{eq18}
    \end{equation}
    \begin{equation}
    P_{\Omega_\textrm{m} - \Omega_\textrm{r}} = g_{\text{pd}} |S_{\text{out},+1}|^2 |S_\textrm{r}|^2
    \label{eq19}
    \end{equation}
    \begin{equation}
    P_{\Omega_\textrm{m} + \Omega_\textrm{r}} = g_{\text{pd}} |S_{\text{out},-1}|^2 |S_\textrm{r}|^2
    \label{eq20}
    \end{equation}
To normalize the measured signals with respect to the optical input
into the device, we measure the carrier transmission at the
vector network analyzer to obtain
    \begin{equation}
    P_\mathrm{in} = g_\mathrm{pd} |S_\mathrm{in}|^2 |S_\mathrm{r}|^2,
    \end{equation}
where $S_\mathrm{in}$ denotes the optical field amplitude inside the
bus waveguide. This quantity can be obtained directly from the transmission
spectrum (assuming low optical loss) when the carrier laser is tuned away from any optical modes,
since at that point the carrier transmission satisfies $S_{\text{out},0} = S_\mathrm{in}$.
By normalizing Eqs.~{\ref{eq18}}–{\ref{eq20}} with $P_\mathrm{in}$, we obtain the expression for the normalized carrier signal and AS and S sideband powers as:
\begin{equation}
P_{\Omega_\textrm{r},\text{norm}} = \dfrac{ |S_{\text{out},0}|^2}{|S_{\text{in}}|^2}
\end{equation}
\begin{equation}
P_{\Omega_\textrm{m} - \Omega_\textrm{r},\text{norm}} = \dfrac{ |S_{\text{out},+1}|^2}{|S_{\text{in}}|^2}
\label{eqS22}
\end{equation}
\begin{equation}
P_{\Omega_\textrm{m} + \Omega_\textrm{r},\text{norm}} = \dfrac{ |S_{\text{out},-1}|^2}{|S_{\text{in}}|^2}
\label{eqS23}
\end{equation}
To see how the normalized sideband power represents the scattering efficiency, we can derive the full expression for Eqs.~{\eqref{eqS22}} and~{\eqref{eqS23}}, according to coupled-mode theory~{\cite{sohn_time-reversal_2018}} as:
\begin{equation}
P_{\Omega_\textrm{m} - \Omega_\textrm{r},\text{norm}} =
 \left|
\frac{i\beta\sqrt{\mbox{$\kappa$}_{\text{ex}1} \mbox{$\kappa$}_{\text{ex}2}}}
{\left( \frac{\mbox{$\kappa$}_2}{2}-i(\omega_l-\omega_2+\Omega_\textrm{m}) \right)
 \left( \frac{\mbox{$\kappa$}_1}{2}-i(\omega_l-\omega_1) \right)}
\right|^2
\end{equation}
\begin{equation}
P_{\Omega_\textrm{m} +\Omega_\textrm{r},\text{norm}} =
 \left|
\frac{i\beta^{*} \sqrt{\mbox{$\kappa$}_{\text{ex}1} \mbox{$\kappa$}_{\text{ex}2}}}
{\left( \frac{\mbox{$\kappa$}_1}{2}-i(\omega_l-\omega_1-\Omega_\textrm{m}) \right)
 \left( \frac{\mbox{$\kappa$}_2}{2}-i(\omega_l-\omega_2) \right)}
\right|^2
\end{equation}
Here $\mbox{$\kappa$}_{1,2}$, $\mbox{$\kappa$}_{\text{ex}1,2}$ and $\omega_{1,2}$ are total loss, extrinsic loss and resonant frequencies of the coupled optical modes (i.e. TE\textsubscript{even} and TE\textsubscript{odd} in the resonator).
Therefore, the normalized scattered powers of the anti-Stokes and Stokes sidebands are proportional to scattering efficiency $|\beta|^2$.
\section*{S7. Robustness and spectral selectivity of non-reciprocal scattering}
As discussed in the main text, the RF index and interaction length of the SWRF electrode were designed to provide an effective RF momentum in the range of $q_{\mathrm{m,SWRF}} = 3300$--$4100~\mathrm{rad/m}$ over $\Omega_\textrm{m}=$ 16--20 GHz. Experimentally, however, the extracted value is $q_{\mathrm{m,SWRF}} = 3240~\mathrm{rad/m}$ at $\Omega_\textrm{m}=16.8$ GHz, which is slightly outside the targeted range. Here we show that robust non-reciprocal scattering is still obtained despite this discrepancy. 
\begin{figure}[H]
  \centering
  \makebox[\textwidth][c]{%
    \includegraphics[width=0.8\textwidth]{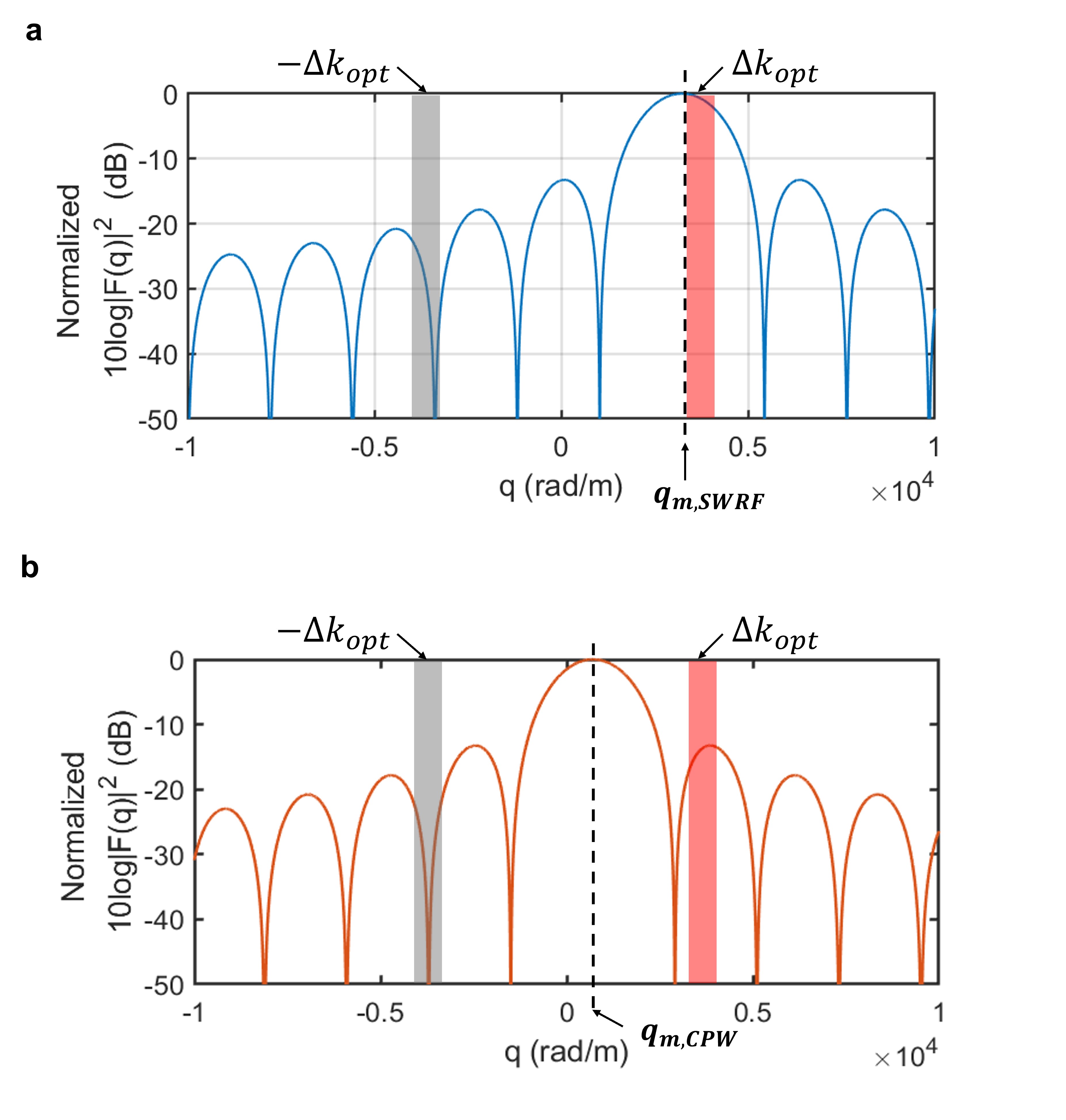}%
  }\caption{Normalized spectral scattering efficiency (dB scale) for the {\textbf{(a)}} SWRF and {\textbf{(b)}} CPW transmission lines for RF drive frequency $\Omega_\textrm{m}=16.8$ GHz and interaction length $\mathcal{L}=2.85$ mm.
  In each case, the efficiency peaks at the modulation momenta $q_{\mathrm{m,SWRF}}$ and $q_{\mathrm{m,CPW}}$. 
  The ranges of $\Delta k_\mathrm{opt}$ and $-\Delta k_\mathrm{opt}$, corresponding to the optical momentum gaps for counter-propagating and co-propagating cases, are highlighted by the red and gray shaded regions. 
}
  \label{figS5}
\end{figure}

Fig.~\ref{figS5}a shows the normalized finite-interaction-length spectral scattering efficiency for $q_{\mathrm{m,SWRF}} = 3240~\mathrm{rad/m}$, plotted on a dB scale. As established in Supplementary \S{S6}, the normalized scattered power $P$ detected in the heterodyne measurement is proportional to $|F(q)|^2$. Consequently, a 20~dB non-reciprocity observed in the measured scattered power between the counter-propagating and co-propagating configurations corresponds directly to a 20~dB asymmetry in the spectral scattering efficiency, i.e., between $|F(q=\Delta k_\mathrm{opt})|^2$ and $|F(q=-\Delta k_\mathrm{opt})|^2$.
From the designed optical dispersion, for a frequency separation of $\Delta \omega_\mathrm{opt} = 16.8~\mathrm{GHz}$, the optical momentum gap 
$\Delta k_\mathrm{opt}$ spans $3340$--$4150~\mathrm{rad/m}$ over the laser tuning range of $1520$--$1570~\mathrm{nm}$. Within this range, the SWRF 
transmission line exhibits robust non-reciprocal contrast between the counter-propagating and co-propagating cases, with a worst case value of $18-19~\mathrm{dB}$.
For comparison, Fig.~\ref{figS5}b shows the normalized finite-interaction-length spectral scattering efficiency of the CPW transmission line. Since $q_{\mathrm{m,CPW}} = 704~\mathrm{rad/m}$, the ranges of $\Delta k_\mathrm{opt}$ and $-\Delta k_\mathrm{opt}$ fall within higher-order lobes of the sinc-shaped spectral response. As a result, although the non-reciprocal contrast depends on the specific choice of finite interaction length, the non-reciprocity achievable with the CPW transmission line over a broad optical wavelength range is intrinsically more limited. In this case, the worst case non-reciprocal contrast between the counter-propagating and co-propagating cases over the laser tuning range of $1520$--$1570~\mathrm{nm}$ is around $5~\mathrm{dB}$, which is substantially smaller than what may be achieved with the SWRF transmission line.

\renewcommand{\refname}{Supplementary References}

\end{bibunit}

\end{document}